\documentclass[aps,prl,twocolumn,superscriptaddress,floatfix,showpacs]{revtex4-1}
\usepackage{epsfig}
\usepackage{dcolumn}
\usepackage{bm}
\usepackage{amssymb}
\usepackage{amsmath}
\usepackage{color}
\usepackage{subfigure}

\pdfoutput=1

\newcommand{\be}{\begin{equation}}
\newcommand{\ee}{\end{equation}}
\newcommand{\bea}{\begin{eqnarray}}
\newcommand{\eea}{\end{eqnarray}}

\begin{document}

\title{Event-by-event hydrodynamics $+$ jet energy loss: A solution to the $R_{AA} \otimes v_2$ puzzle}

\author{Jacquelyn Noronha-Hostler}
\affiliation{Department of Physics, University of Houston, Houston TX 77204, USA}
\author{Barbara Betz}
\affiliation{Institut f\"ur Theoretische Physik, Goethe Universit\"at, Frankfurt, Germany}
\author{Jorge Noronha}
\affiliation{Instituto de F\'{\i}sica, Universidade de S\~{a}o Paulo, C.P. 66318,
05315-970 S\~{a}o Paulo, SP, Brazil}
\author{Miklos Gyulassy}
\affiliation{Nuclear Science Division, Lawrence Berkeley National Laboratory, Berkeley, CA 94720, USA}
\affiliation{Pupin Lab MS-5202, Department of Physics, Columbia University, New York, NY 10027, USA}
\affiliation{Institute of Particle Physics, Central China Normal University, Wuhan, China}

\date{\today}

\begin{abstract}
High $p_T > 10$ GeV elliptic flow, which is experimentally measured via the correlation between soft and hard hadrons, receives competing contributions from event-by-event fluctuations of the low $p_T$ elliptic flow and event plane angle fluctuations in the soft sector. In this paper, a proper account of these event-by-event fluctuations in the soft sector, modeled via viscous hydrodynamics, is combined with a jet energy loss model to reveal that the positive contribution from low $p_T$ $v_2$ fluctuations overwhelms the negative contributions from event plane fluctuations. This leads to an enhancement of high $p_T > 10$ GeV elliptic flow in comparison to previous calculations and provides a natural solution to the decade long high $p_T$ $R_{AA} \otimes v_2$ puzzle. We also present the first theoretical calculation of high $p_T$ $v_3$, which is shown to be compatible with current LHC data. Furthermore, we discuss how short wavelength jet-medium physics can be deconvoluted from the physics of soft, bulk event-by-event flow observables using event shape engineering techniques.

\end{abstract}

\maketitle

\noindent \textsl{1. Introduction.} Inspired by Bjorken's original jet quenching idea \cite{bjorken}, the energy loss experienced by fast moving partons in the Quark-Gluon Plasma (QGP) formed in heavy ion collisions has been studied using the nuclear modification factor (at mid-rapidity) $R_{AA}(p_T,\phi) = \frac{dN_{AA}/dp_T d\phi}{N_{\rm coll}\,dN_{pp}/dp_T}$, where $dN_{AA}/dp_T$ is the spectrum of the corresponding particle species (e.g., pions) in 
AA collisions, $dN_{pp}/dp_T$ is the corresponding proton-proton yield, $\phi$ is the azimuthal angle in the plane transverse to the beam direction, and $N_{\rm coll}$ is the total number of binary collisions \cite{Miller:2007ri}. The azimuthally averaged version of this quantity, $R_{AA}(p_T) =\frac{1}{2\pi} \int_0^{2\pi}d\phi\, R_{AA}(p_T,\phi)$, was predicted \cite{Gyulassy:1990ye,Wang:1991hta,Wang:1991xy,Vitev:2002pf} and later experimentally observed at RHIC 
\cite{Adcox:2001jp,Adler:2002xw,Adler:2003qi,Adams:2003kv,Adams:2003im} to strongly depend on geometrical 
control parameters that involve length scales of the order of the radius of a large nucleus (e.g., 
$\sim 7.5$ fm for Au) such as the centrality (multiplicity) of the collisions. This program has been successfully extended to LHC energies \cite{Spousta:2013aaa}, whose central collisions produce a hotter, twice as dense QGP than the one formed at RHIC's top energies \cite{Horowitz:2011gd}. The combination of RHIC and LHC data has been instrumental to determine how jets couple with the evolving medium \cite{Armesto:2011ht,Burke:2013yra} and to motivate new theoretical and phenomenological studies \cite{Liao:2008dk,MehtarTani:2010ma,Noronha:2010zc,Jia:2011pi,Ovanesyan:2011xy,Renk:2011qi,Buzzatti:2011vt,Young:2011ug,Qin:2012fua,Zhang:2012ha,D'Eramo:2012jh,Blaizot:2013hx,Molnar:2013eqa,Djordjevic:2013pba,Ficnar:2013qxa,Li:2014hja,Uphoff:2014cba,Xu:2014ica,Andrade:2014swa,Casalderrey-Solana:2014bpa,Betz:2014cza,Xu:2014tda,Ayala:2015jaa,Rougemont:2015wca,Xu:2015bbz,Ghiglieri:2015ala,Casalderrey-Solana:2015tas,Betz:2015mlf}.  

It was recognized early on \cite{Wang:2000fq,Gyulassy:2000gk,Shuryak:2001me} that the azimuthal anisotropy of high $p_T$ hadrons encoded in $R_{AA}(p_T,\phi)$ was a powerful tool to study the energy loss and the path length dependence of hard partons in the QGP. The anisotropic flow coefficients associated with $R_{AA}(p_T,\phi)$ can be computed from its Fourier series
\be
\frac{R_{AA}(p_T,\phi)}{R_{AA}(p_T)} =  1  + 2\sum_{n=1}^\infty v_n^{hard}(p_T) \cos\left[n\phi - n\psi_n^{hard}(p_T)  \right]
\ee
where
\be
\label{definevnhard}
v_n^{hard}(p_T) = \frac{\frac{1}{2\pi}\int_0^{2\pi}d\phi\,\cos\left[n\phi-n\psi_n^{hard}(p_T)\right]\,R_{AA}(p_T,\phi)}{R_{AA}(p_T)}
\ee
and $\psi_n^{hard}(p_T) = \frac{1}{n}\tan^{-1}\left(\frac{\int_0^{2\pi}d\phi\,\sin\left(n\phi\right)\,R_{AA}(p_T,\phi)}{\int_0^{2\pi}d\phi\,\cos\left(n\phi\right)\,R_{AA}(p_T,\phi)}\right)$. While the azimuthally averaged $R_{AA}(p_T)$ can be described by many different models, it has proven to be a challenge in the field (see discussion in Refs.\ \cite{Betz:2014cza,Xu:2014tda}) to obtain a simultaneous description of $R_{AA}(p_T)$ and high $p_T$ elliptic flow. Model calculations typically give a small $v_2^{hard}(p_T)$ that is incompatible with elliptic flow data. 


An important detail that has been overlooked so far in model calculations is that the theoretical $v_2^{hard}(p_T)$ is not the appropriate quantity to be compared with experimental data. In fact, the experimental  high $p_T> 10$ GeV flow coefficients $v^{exp}_n(p_T)$ are measured via the \emph{correlation} between soft and hard hadrons in a given centrality class,  emphasized in \cite{Luzum:2013yya,Paquet:2015lta}
\begin{equation}\label{eqn:vncor}
v^{exp}_n(p_T)=\frac{\langle v^{soft}_n\,v_n^{hard}(p_T)\cos\left[n\left(\psi^{soft}_n-\psi^{hard}_n(p_T)\right]\right) \rangle}{\sqrt{\left\langle \left(v^{soft}_n\right)^{2}\right\rangle}},
\end{equation}
where $v_n^{soft}$, $\psi_n^{soft}$ are the integrated soft flow harmonic and the corresponding event plane angle \cite{eventplane} for all charged particles with $p_T \lesssim 3$ GeV and where $\langle \ldots \rangle $ experimentally denotes an average over all events. The idealized limit, $v_2(p_T \gtrsim\, 10\, {\rm GeV}) \sim v_2^{hard}(p_T)$, considered in previous model calculations, is not realistic since it neglects event-by-event fluctuations of the bulk geometry. Thus, a description of high $p_T$ anisotropic flow $v_n(p_T)$ necessitates  modeling of both the soft and hard sectors of heavy ion collisions.

Our understanding of the bulk (soft) properties of the evolving medium has progressed 
immensely since the first event-by-event hydrodynamic simulations were carried out more than a 
decade ago \cite{Gyulassy:1996br,Aguiar:2001ac,Socolowski:2004hw,Andrade:2006yh}. It is now understood 
\cite{Alver:2010gr} that the initial spatial anisotropies present in the early stages of 
nucleus-nucleus collisions are converted to final stage momentum anisotropies (i.e., anisotropic 
flow) in a way that is consistent with viscous relativistic hydrodynamic simulations performed on 
an event-by-event basis (see, e.g., the review \cite{Luzum:2013yya}). Furthermore, once the full 
information regarding the soft event-by-event $v_n$ distributions became available \cite{Aad:2013xma} 
at the LHC, powerful constraints on the initial conditions of the hydrodynamic modeling of the 
QGP have been obtained \cite{Gale:2012rq,Niemi:2012aj,Renk:2014jja,Niemi:2015qia,Noronha-Hostler:2015dbi}. 
Nevertheless, these recent advances regarding event-by-event medium fluctuations have not yet been incorporated in theoretical studies of high $p_T$ observables.  

\begin{figure}[ht]
\centering
\includegraphics[width=0.4\textwidth]{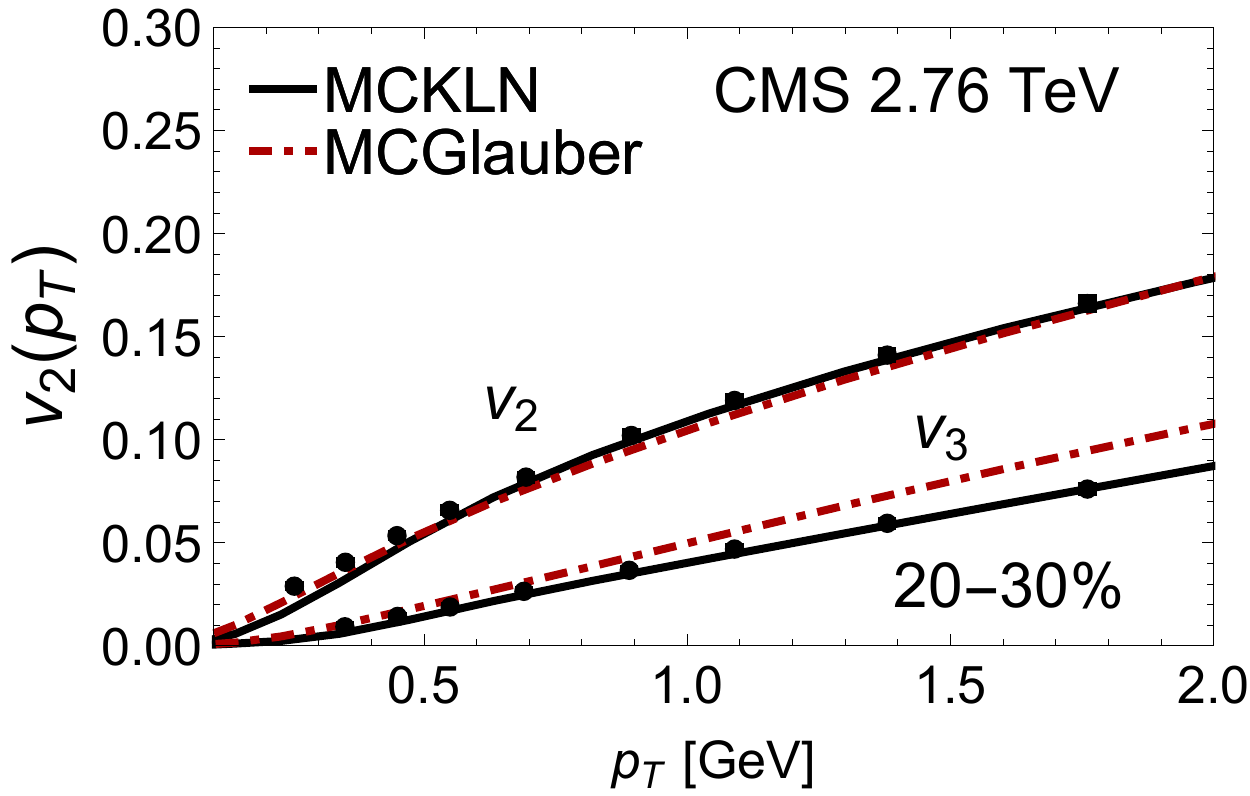}
\caption{(Color online) Model comparison to CMS 
data \cite{Chatrchyan:2013kba} for low $p_T<2$ GeV $v_2(p_T)$ and $v_3(p_T)$ of charged particles.}
\label{fig:distrov2v3}
\end{figure}

In this paper we show that the long standing high $p_T$ $R_{AA} \otimes v_2$ puzzle can be naturally solved by computing high $p_T$ elliptic flow using its experimental definition in \eqref{eqn:vncor} that takes into account the effects of event-by-event fluctuations needed for a realistic description of the QGP. The soft sector is modeled via event-by-event hydrodynamic simulations performed using the v-USPhydro code \cite{Noronha-Hostler:2013gga,Noronha-Hostler:2014dqa,Noronha-Hostler:2015coa,Noronha-Hostler:2015wft} while the hard sector is described using the energy loss framework developed in \cite{Betz:2011tu,Betz:2012qq,Betz:2014cza}. We show that the positive contribution from low $p_T$ $v_2$ fluctuations overwhelms the small, negative contributions from event plane fluctuations, which in turn leads to an overall enhancement of high $p_T$ elliptic flow in comparison to previous calculations. The inclusion of initial state fluctuations followed by viscous hydrodynamics allowed us to perform the first theoretical calculation of high $p_T$ $v_3$, which is shown to be compatible with current LHC data.

\noindent \textsl{2. Hydrodynamic evolution.} The expanding QGP is modeled through 
event-by-event simulations performed using the 2+1 (i.e., boost invariant) viscous relativistic 
hydrodynamics code called v-USPhydro \cite{Noronha-Hostler:2013gga,Noronha-Hostler:2014dqa,Noronha-Hostler:2015coa,Noronha-Hostler:2015wft}. 
v-USPhydro accurately \cite{info} solves the energy-momentum conservation equations and the equations 
of motion for the dissipative currents using the Lagrangian formulation of hydrodynamics encoded in the 
Smoothed Particle Hydrodynamics (SPH) algorithm \cite{SPH,Aguiar:2000hw}. Information about four transport coefficients is required: the temperature-dependent shear and bulk viscosities, $\eta$ and 
$\zeta$, and their respective relaxation time coefficients, $\tau_\pi$ and $\tau_\Pi$ (other 2nd order 
transport coefficients \cite{Finazzo:2014cna} are not yet taken into account). For simplicity, effects 
from the temperature dependence of $\eta/s$ in the hadronic 
\cite{NoronhaHostler:2008ju,Demir:2008tr,NoronhaHostler:2012ug} or in the QGP phase \cite{Niemi:2011ix} 
are neglected here and, thus, we set $\eta/s$ to be a constant. Also, in this first study bulk viscosity \cite{Noronha-Hostler:2013gga,Noronha-Hostler:2014dqa,Ryu:2015vwa} is set to zero. The initial time 
for all the hydrodynamic simulations was $0.6$ fm and we used the lattice-based equation of 
state EOS S95n-v1 \cite{EOS} and an isothermal Cooper-Frye \cite{CF} freezeout with freeze-out 
temperature $T_F = 120$ MeV for MCKLN and $T_F = 130$ for MCGlauber. Particle decays are included 
(with hadronic resonances with masses up to 1.7 GeV) via an adapted version of the AZHYDRO code \cite{azhydro}.

MCGlauber and MCKLN initial conditions \cite{Drescher:2006ca} for the mid-central $20-30\%$ centrality class of $\sqrt{s}=2.76$ 
TeV Pb+Pb collisions at the LHC that provide a good description of this eccentricity data (our results agree with Ref.\ \cite{Aad:2013xma}). 
Our results for the low $p_T$ $v_2$ and $v_3$ for all charged particles (here averaged over 150 events) and the comparison to CMS data 
\cite{Chatrchyan:2013kba} ($\eta/s=0.08$ in MCGlauber and $\eta/s=0.11$ in MCKLN) determined 
using the event plane method \cite{eventplane} are shown in Fig.\ \ref{fig:distrov2v3}. While for $v_2(p_T)$ both types of initial 
conditions give a reasonable description of the low $p_T$ data,
due to $v_3(p_T)$ MCKLN is slightly favored over MCGlauber in our calculations.


\noindent \textsl{3. Energy loss model.} The nuclear modification factor 
and the high $p_T$ azimuthal anisotropies are studied here using the BBMG jet-energy loss model 
developed in \cite{Betz:2011tu,Betz:2012qq,Betz:2014cza} supplemented with the energy density 
and flow profiles obtained from the event-by-event viscous hydrodynamic simulations for the soft 
sector described above. In this model, the energy loss per unit length experienced by a fast 
moving parton in the plasma, $dE/dL$, is modeled as
\be
\frac{dE}{dL} = -\kappa\,
E^a(L) \,L^z\, T^c \,\zeta_q\, \Gamma_{\rm flow}
\ee 
where $\kappa$ is the jet-medium coupling \cite{Betz:2014cza}, $T$ is the local temperature 
field along the jet trajectory with $c=2+z-a$, $\zeta_q$ describes energy loss fluctuations 
\cite{Betz:2014cza}, 
$\Gamma_{\rm flow} = \Gamma_{\rm f}= \gamma \left[1-v \cos\left(\phi_{\rm jet}-\phi_{\rm flow}  \right)\right]$ 
is the flow factor defined using the local flow velocities of the medium $\vec{u}=\gamma \vec{v}$ 
(with $\gamma = 1/\sqrt{1-\vec{v}^{\,2}}$) \cite{Armesto:2004vz,Renk:2005ta,Baier:2006pt,Liu:2006he}, 
$\phi_{\rm jet}$ is the angle defined by the propagating jet in the transverse plane, and 
$\phi_{\rm flow}$ is the local azimuthal angle of the hydrodynamic flow. In this framework 
the dependence of the energy loss rate with the jet energy $E$, path length $L$, temperature $T$, 
and energy loss fluctuations $\zeta_q$ is characterized by the parameters $(a, z, c, q)$.

\begin{figure*}[ht]
\centering
\includegraphics[width=0.8\textwidth]{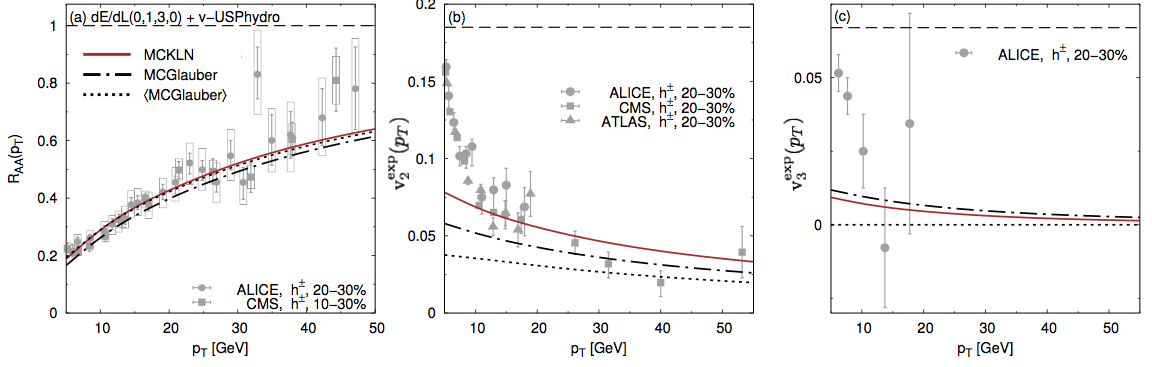}
\caption{(Color online) Model calculations for (a) 
$\pi^0$ $R_{AA}(p_T)$, (b) $v_2^{exp}(p_T)$, (c) $v_3^{exp}(p_T)$ in 
mid-central $\sqrt{s}=2.76$ TeV Pb+Pb collisions at the LHC. MCKLN initial conditions are shown in solid red while the dotted-dashed black line is for MCGlauber. The black dotted line $\langle{\rm MCGlauber}\rangle$ corresponds to results obtained neglecting any initial state fluctuations of the soft bulk background by evolving only an event averaged smoothed initial Glauber geometry. The experimental data are taken from Refs.\ \cite{ALICE_RAA,CMS_RAA,ALICE_v2_v3,CMS_v2,ATLAS_v2}. }
\label{fig:RAAv2v3}
\end{figure*}

We focus in this Letter 
on the ``pQCD-scenario" 
discussed in detail in Ref.\ \cite{Betz:2014cza} where $(a=0,z=1,c=3,q=0)$, i.e., $dE/dL \sim L$. Other dependences on the path length will be discussed elsewhere. The jets are distributed according to the given transverse profile for the medium given by v-USPhydro. The jet path
$\vec{x}(L)=\vec{x}_0+\hat{n}(\phi_{\rm jet})L$ from a production point $\vec{x}_0$ is perpendicular to the beam axis
and moves in the direction given by $\phi_{\rm jet}$. All jet production points with local temperature above
$160$ MeV are taken into account. In this study we 
used the KKP pion fragmentation functions \cite{Kniehl:2000hk}, which have been tested against 
RHIC and LHC data \cite{Simon:2006xt}. 
\begin{figure}[ht]
\centering
\includegraphics[width=0.3\textwidth]{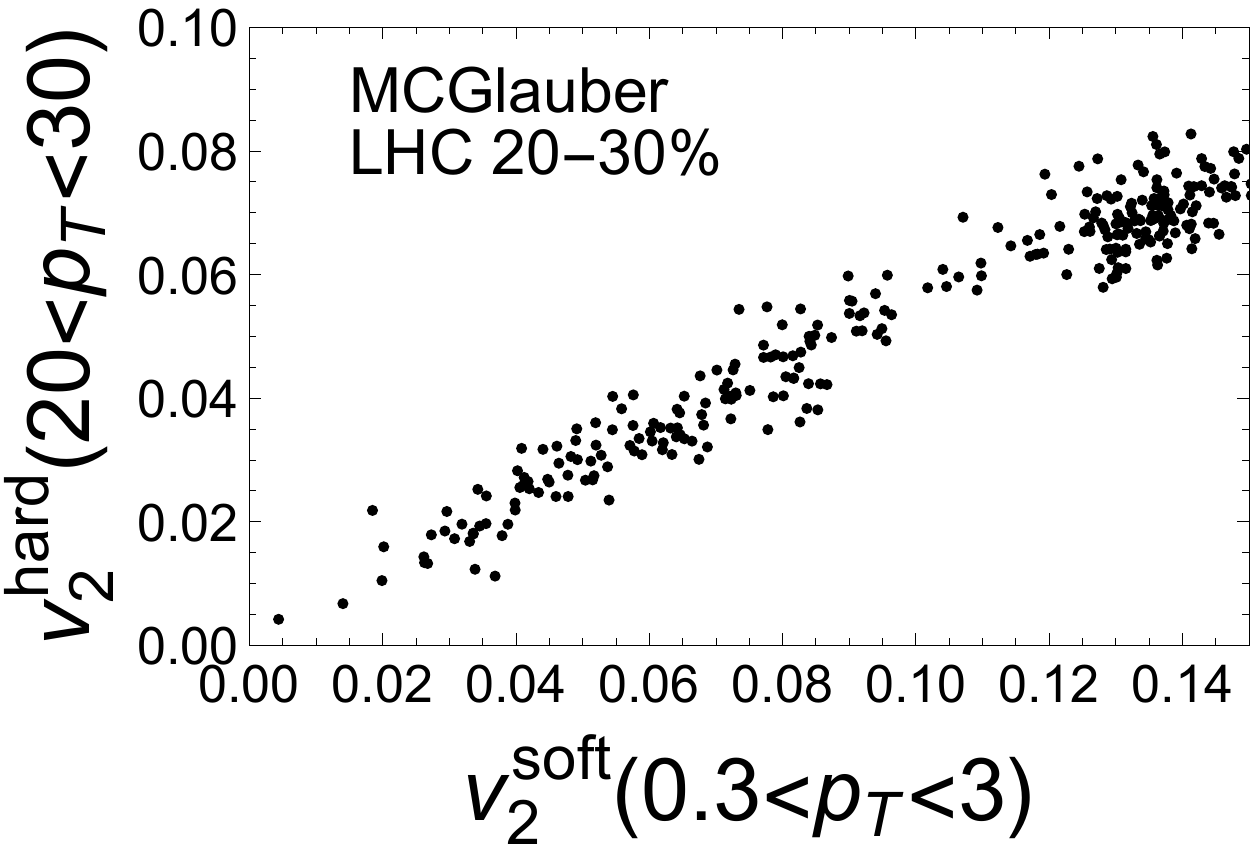}
\caption{Event-by-event correlation between $v_2^{soft}$ (computed via viscous hydrodynamics) and $v_2^{hard}$ (defined in Eq.\ \eqref{definevnhard}). The approximate linear correlation indicates that initial state fluctuations, which determine $v_2^{soft}$, also directly affect the 2nd harmonic of $R_{AA}(p_T,\phi)$. }
\label{fig:linearresponse}
\end{figure}

\noindent \textsl{4. Results for $R_{AA}$ and $v_n$ at high $p_T$.} The $\kappa$ parameter in the BBMG energy loss model is completely fixed (for each type of initial 
conditions) by matching the computed $\pi^0$ $R_{AA}(p_T = 10 \,{\rm GeV})$ to all-charged data in 0-5\% most 
central LHC collisions (initial comparisons between $\pi^0$ and all charged for $v_{_{2}}$ of high-$p_{_{T}}$ are very similiar \cite{Hamed:2014eja}). Our results for $\pi^0$ $R_{AA}(p_T)$ in mid-central Pb+Pb $\sqrt{s}=2.76$ TeV collisions at the LHC and the experimental data \cite{ALICE_RAA,CMS_RAA} are shown in Fig.\ 
\ref{fig:RAAv2v3} (a). The solid red line corresponds to the results computed using 
the hydrodynamic evolution based on the MCKLN initial condition while 
the dotted-dashed black curve denotes results computed using MCGlauber initial conditions. The black dotted line  corresponds to results obtained neglecting any initial state fluctuations of the soft background by evolving only an event averaged, smoothed initial Glauber geometry. While as expected $R_{AA}(p_T)$ is robust with respect to the inclusion of event-by-event fluctuations, the same cannot be said about the anisotropic flow coefficients $v_2(p_T)$ and $v_3(p_T)$, computed using \eqref{eqn:vncor}, and shown in Fig.\ 
\ref{fig:RAAv2v3} (b)  and (c). By comparing the dotted and the dashed-dotted curves (both computed using MCGlauber) one can see that the inclusion of event-by-event fluctuations significantly enhances $v_2(p_T)$ and gives a nonzero $v_3(p_T)$. The larger average eccentricity in MCKLN gives larger $v_2(p_T)$ (solid red line) in comparison to MCGlauber results but the opposite is found for $v_3(p_T)$, which is in accordance with the fact that MCKLN gives smaller values for $\varepsilon_3$ in comparison to MCGlauber's (see, for instance, \cite{Noronha-Hostler:2015coa}). One can see in Fig.\ \ref{fig:RAAv2v3} that a reasonable simultaneous description of $R_{AA}$ and $v_2^{exp}(p_T)$ data can be obtained in this approach. Also, $v_3^{exp}(p_T)$ is in the ballpark of current data uncertainties, which indicates that the initial state fluctuations that generate $v_3$ in the soft sector \cite{Alver:2010gr} are also responsible for triangular flow at high $p_T$.

There is a simple way to understand why event-by-event fluctuations increase $v_2^{exp}(p_T)$ in comparison to event averaged calculations. First, we observe that the 2nd flow harmonic $v_2^{hard}(p_T)$ defined in \eqref{definevnhard} fluctuates event-by-event and it is, to a good approximation, \emph{linearly correlated} with its soft counterpart. Indeed, we show in Fig.\ \ref{fig:linearresponse} that the integrated $v_2^{hard}(20 < p_T< 30\, {\rm GeV})$ is linearly correlated with $v_2^{soft}(0.3<p_T<3 \,{\rm GeV})$ on an event-by-event basis (which is similar to what is observed in the data \cite{Aad:2015lwa}). This shows that the initial state geometrical fluctuations responsible for $v_2^{soft}$ lead to fluctuations in the path length of the jet in the medium. Using this linear correlation and that on average $\psi_2^{hard}(p_T)$ is aligned with $\psi_2^{soft}$ \cite{Jia:2012ez}, one can see that the effects of \emph{small} fluctuations (kept up to quadratic order) on the soft-hard correlation in \eqref{eqn:vncor} are
\be
\frac{v_2^{exp}(p_T)}{\left\langle v_2^{hard}(p_T) \right\rangle} \simeq   1 + \frac{1}{2}\left\langle \left(\frac{\delta v_2^{soft}}{\langle v_2^{soft} \rangle} \right)^2  \right\rangle  -2 \left\langle  \left(\delta \psi_2(p_T)\right)^2 \right \rangle \,,
\ee
where $\delta \psi_2(p_T) = \delta (\psi_2^{hard}(p_T) - \psi_2^{soft})$. Event-by-event fluctuations enhance experimental elliptic flow because the positive contribution from fluctuations of the soft $\delta v_2^{soft}$ overwhelms the small, negative contribution from event plane misalignment. We found that $\langle\cos\left[2\delta \psi_2\right]\rangle\sim 0.99$ while $\langle\cos\left[3\delta \psi_3\right]\rangle\sim 0.33$ in our simulation. It would be interesting to test if there is an analog of the mapping of the eccentricities onto the final high $p_T$ flow harmonics that also mirrors the soft correlation seen at low $p_T$ \cite{Gardim:2011xv,Gardim:2014tya,Bhalerao:2011bp}.

The new approach pursued, which combines event-by-event hydrodynamics with jet energy loss, can be used to exploit initial geometrical shape fluctuations as an additional experimental control knob besides centrality by measuring $v_n^{exp}(p_T;C^{soft})$ in different soft bulk multiplicity and eccentricity subclasses, (e.g, ${\cal C}^{soft}=\{
\Delta N_{tracks}/N_{max}=20-30\%,\;\Delta v_2^{soft}/\langle v^{soft}_{2}\rangle =10\% \}$), through Eq.\ \eqref{eqn:vncor}. Thus, we propose to couple event shape engineering \cite{Schukraft:2012ah,Aad:2015lwa,Adam:2015eta} 
in the low $p_T$ soft sector with jet quenching observables in subclasses of spatially anisotropic events in the same centrality class. By taking a very narrow window of events near the average $\langle v_2^{soft}\rangle$ of the wide $v_2^{soft}$ distribution, one can systematically reduce the effects of fluctuations in the soft sector to approach the ideal theoretical limit $v_2^{exp}(p_T) \to v_2^{hard}(p_T)$ at high $p_T$. This would allow one for the first time to deconvolute the short wavelength jet-medium physics contained in the azimuthal dependence $R_{AA}(p_T,\phi)$ from the physics of soft, bulk event-by-event flow observables. This type of soft-hard event engineering (SHEE) would also allow for novel studies of the path length dependence of energy loss in highly anisotropic media (associated with the events at the tail of the $v_2^{soft}$ distribution). A dedicated study about SHEE will be presented elsewhere.


\noindent \textsl{5. Conclusions.} In this paper, event-by-event fluctuations in the soft sector modeled via viscous hydrodynamics were combined with a jet energy loss model to solve the decade long high $p_T$ $R_{AA} \otimes v_2$ puzzle in ultrarelativistic heavy ion collisions. A crucial point to this study was the realization that the experimentally measured high $p_T$ azimuthal coefficients are currently defined via a correlation between soft and hard particles over many events, see \eqref{eqn:vncor}. Therefore, these observables inherit the well-known geometrical fluctuations of the soft sector (see the linear correlation in Fig.\ \ref{fig:linearresponse}) and the failure of previous model calculations to simultaneously describe $R_{AA}(p_T)$ and $v_2(p_T)$ stem not from the lack of some non-perturbative source but rather from unrealistic assumptions for the evolving medium (event averaged instead of a realistic event-by-event hydrodynamic evolution). We showed that the positive contribution from low $p_T$ $v_2^{soft}$ fluctuations overwhelms the negative contributions from event plane fluctuations and this leads to an overall enhancement of high $p_T > 10$ GeV elliptic flow in comparison to previous calculations. Also, the inclusion of event-by-event fluctuations allowed us to compute for the first time high $p_T$ $v_3^{exp}$, which displayed a reasonable agreement with data (given current uncertainties). 

Our study paves the way for a new era in high $p_T$ physics in which state-of-the-art experimental event shape engineering techniques may be used to deconvolute the (strongly coupled) bulk evolution of the QGP from the perturbative QCD energy loss physics that determines the azimuthal anisotropy of jet quenching phenomena. We encourage experimentalists to check the features of low $p_T$ $v_n^{soft}$ distributions, cumulants, and $\langle p_T \rangle$ of events with triggered high $p_T$ particles in the context of event shape engineering.  Future experimental SHEE studies at RHIC and LHC varying both centrality and soft eccentricity subclasses may help to discriminate better between other alternative combinations of soft-hard dynamical models of high energy nuclear collisions.

\noindent \textsl{Acknowledgments.} We thank G.~S.~Denicol, M.~Luzum, S.~Mohapatra, A.~Timmins, W.~Li, and J.~Liao for discussions. JNH was supported by the University of Houston. BB acknowledges support from the Bundesministerium f\"ur Bildung und Forschung and the Helmholtz
International Center for FAIR. JNH and MG were supported in part by US-DOE Nuclear Science Grant No. DE-FG02-93ER40764. MG also acknowledges partial support by the IOPP, CCNU, Wuhan, China. JN thanks the University 
of Houston for its hospitality and Funda\c c\~ao de Amparo \`a Pesquisa do Estado de 
S\~ao Paulo (FAPESP) and Conselho Nacional de Desenvolvimento Cient\'ifico e Tecnol\'ogico (CNPq) 
for support.

\end{document}